# De spanning tussen het non-discriminatierecht en het gegevensbeschermingsrecht: heeft de AVG een nieuwe uitzondering nodig om discriminatie door kunstmatige intelligentie tegen te gaan?

Marvin van Bekkum & Frederik Zuiderveen Borgesius[1]

marvin.vanbekkum@ru.nl & frederikzb@cs.ru.nl

Organisaties kunnen kunstmatige intelligentie (*artificial intelligence, AI*) gebruiken om beslissingen te nemen over mensen, bijvoorbeeld om de beste kandidaten te selecteren uit vele sollicitatiebrieven. Maar AI kan discriminerende effecten hebben. Een AI-systeem zou bijvoorbeeld ten onrechte sollicitaties van mensen met een bepaalde etniciteit kunnen afwijzen, terwijl de organisatie dat niet zo bedoelde. Als de organisatie wil controleren of zijn AI-systeem discrimineert op etniciteit, stuit zij op een probleem: de organisatie kent vaak de etniciteit van haar sollicitanten niet. In beginsel verbiedt de AVG het gebruik van bepaalde categorieën gevoelige persoonsgegevens, waaronder etniciteit. Dit artikel bespreekt de problematiek rond het AVG-verbod op het verzamelen van gevoelige persoonsgegevens met het doel om AI-systemen te controleren op discriminatie. We brengen ook in kaart welke voor- en tegenargumenten er zijn voor het creëren van een nieuwe uitzondering op het AVG-verbod om zulke gevoelige gegevens te gebruiken om discriminatie door AI-systemen tegen te gaan.

---

[1] Mr. drs. M.S.L. van Bekkum onderzoekt als promovendus de discriminatierisico's van AI-systemen in de verzekeringssector. Prof. dr. F.J. Zuiderveen Borgesius is hoogleraar ICT en recht. Beiden zijn verbonden aan de iHub, de interdisciplinary research hub on digitalization and society, Radboud Universiteit. Dit artikel is gebaseerd op: Van Bekkum & Zuiderveen Borgesius, 'Using sensitive data to prevent discrimination by artificial intelligence: Does the GDPR need a new exception?', Computer Law & Security Review, Volume 48, April 2023.

## 1   Inleiding

Stel dat een organisatie een AI-systeem gebruikt om de beste kandidaten te selecteren uit honderden sollicitatiebrieven. De organisatie wil onbedoelde discriminatie voorkomen. Ze wil checken of het AI-systeem ten onrechte alle brieven afkeurt van, bijvoorbeeld, mensen met een immigratieachtergrond. Om te controleren of zijn AI-systeem mensen met een bepaalde etniciteit benadeelt, moet de organisatie de etniciteit van haar sollicitanten kennen. Maar de Algemene verordening gegevensbescherming (AVG) bevat een verbod (met uitzonderingen) op het gebruik van 'bijzondere persoonsgegevens' (artikel 9). Bijzondere persoonsgegevens zijn onder meer gegevens over etniciteit, religie, en seksuele voorkeur.

Dit artikel bespreekt twee vragen. (i) Belemmeren de regels van de AVG over bijzondere persoonsgegevens het tegengaan van discriminatie door AI-systemen? (ii) Wat zijn de argumenten voor en tegen het creëren van een nieuwe uitzondering op het AVG-verbod op het gebruik van bijzondere persoonsgegevens om discriminatie door AI-systemen tegen te gaan?

Een uitzondering op het AVG-verbod op het gebruik van bijzondere persoonsgegevens zou kunnen worden opgenomen in de AVG, of in een andere wet, nationaal of op EU-niveau. Wij richten ons alleen op discriminatie met betrekking tot bepaalde beschermde gronden in de non-discriminatierichtlijnen van de EU, namelijk etniciteit, godsdienst of overtuiging, handicap en seksuele geaardheid.[2] In Nederland zijn die richtlijnen grotendeels geïmplementeerd in de Algemene wet gelijke behandeling (Awgb).

Dit artikel kan relevant zijn voor ander andere juristen en computerwetenschappers in de academie en de praktijk, en voor beleidsmakers. De meeste bestaande literatuur over het gebruik van bijzondere persoonsgegevens voor non-discriminatiedoeleinden is versnipperd over disciplines en sub-disciplines. Wij combineren inzichten uit het non-discriminatierecht enerzijds met die uit het privacy- en gegevensbeschermingsrecht anderzijds. We houden ook rekening met inzichten uit de informatica en AI.

---

[2] Ras of etnische afstamming: Richtlijn 2000/43/EG PB L 180/22. Godsdienst of overtuiging, handicap, leeftijd of seksuele geaardheid in de arbeidscontext: Richtlijn 2000/78/EG PB L 303/16. Geslacht in de context van de levering van goederen en diensten: Richtlijn 2004/113/EG 2004 PB L 373/37. Geslacht in de arbeidscontext: Richtlijn 2006/54/EG 2006 PB L 204/23. Leeftijd en geslacht zijn geen bijzondere persoonsgegevens, hoewel het beschermde discriminatiegronden zijn. Zie artikel 9 lid 1 AVG.

We laten zien dat de AVG in de meeste omstandigheden niet toestaat dat een organisatie bijzondere persoonsgegevens gebruikt voor het tegengaan van AI-discriminatie. Wij zijn het dus niet eens met sommige non-discriminatierecht-juristen die lijken te suggereren dat de AVG dergelijk gegevensgebruik toestaat.[3] De AVG staat de EU en de lidstaten toe om een uitzondering aan te nemen, maar dat is tot nu toe niet gebeurd. De EU heeft wel zo'n uitzondering voorgesteld in het voorstel voor een AI-verordening; die bespreken we in paragraaf 7.

Het is waarschijnlijk, maar niet zeker, dat het voorstel voor de AI-verordening wordt aangenomen. Ook als de AI-verordening wordt aangenomen, dan is niet zeker of de uitzondering op het AVG-verbod nog voorkomt in de definitieve tekst. Ons artikel kan helpen de voorgestelde uitzondering beter te begrijpen. En als de AI-verordening niet wordt aangenomen, of wordt aangenomen zonder de uitzondering, dan zou de Nederlandse wetgever kunnen overwegen zo'n uitzondering aan te nemen.

Het ontwikkelen van niet-discriminerende AI-systemen en het controleren van bestaande AI-systemen staat in Nederland hoog op de agenda. Bouwers van AI-systemen moeten gedurende het hele ontwikkelproces rekening houden met non-discriminatienormen (*non-discriminatie by design*).[4] Het controleren van AI-systemen op discriminatie is bovendien een belangrijk onderdeel van een impact assessment die tot doel heeft de risico's van AI-systemen in kaart te brengen, zoals de 'Impact Assessment Mensenrechten en Algoritmes'.[5] Een organisatie die zulke procedures volgt zal rekening moeten houden met het AVG-verbod en de discussie daaromheen.

---

[3] Zie onder 5.2.
[4] Zie Ministerie van Binnenlandse Zaken en Koninkrijksrelaties, Handreiking non-discriminatie Artificial Intelligence (AI), 2022, https://www.rijksoverheid.nl/documenten/rapporten/2022/12/05/handreiking-non-discriminatie-artificial-intelligence-ai, p. 28 & 30. Dit rapport is gebaseerd op Van der Sloot e.a., *Handreiking voor niet discriminerende algoritmes*, TILT 2021, https://www.tilburguniversity.edu/nl/over/schools/law/departementen/tilt/onderzoek/handreiking.
[5] Ministerie van Binnenlandse Zaken en Koninkrijksrelaties, Impact Assessment, *Mensenrechten en Algoritmes*, Ministerie van Binnenlandse Zaken en Koninkrijksrelaties 2022, https://www.rijksoverheid.nl/documenten/rapporten/2021/02/25/impact-assessment-mensenrechten-en-algoritmes.

Het artikel is als volgt opgebouwd. In paragraaf 2 bespreken we hoe AI-systemen kunnen discrimineren op basis van etniciteit en vergelijkbare kenmerken. In paragraaf 3 introduceren we de non-discriminatiewetgeving. In paragraaf 4 en 5 analyseren we de AVG-regels voor bijzondere persoonsgegevens en laten we zien dat die regels het gebruik van bijzondere persoonsgegevens belemmeren. In paragraaf 6 analyseren we de argumenten voor en tegen de invoering van een nieuwe uitzondering op het AVG-verbod voor de controle van AI-systemen. Paragraaf 7 bespreekt mogelijke waarborgen die met een nieuwe uitzondering gepaard kunnen gaan, en paragraaf 8 sluit af.

## 2    AI-systemen en discriminatie

Artificial Intelligence kan worden omschreven als 'het vermogen van een computer om gegevens te verwerken waarbij zo veel mogelijk wordt geprobeerd het menselijk denken na te bootsen.'[6] Enkele voorbeelden zijn computergestuurde beeldherkenning, spraakherkenning, besluitvorming en vertaling tussen talen. Dit artikel bespreekt AI-systemen die beslissingen nemen die ernstige gevolgen kunnen hebben voor mensen. Een bank zou bijvoorbeeld een AI-systeem kunnen gebruiken om te beslissen of een klant een hypotheek krijgt of niet. In dit artikel richten we ons op het risico dat AI-systemen discrimineren tegen mensen met bepaalde beschermde kenmerken, zoals etniciteit, religie, of seksuele geaardheid.

Discriminerende *training data* zijn een van de belangrijkste bronnen van discriminatie door AI-systemen.[7] Zo zijn sommige AI-gestuurde gezichtsherkenningssystemen getraind op foto's van witte mensen. Zulke systemen kunnen dan slecht zijn in het herkennen van mensen met een andere huidskleur.[8]

---

[6] Dikke Van Dale 2023, definitie 'Kunstmatige Intelligentie', https://www.vandale.nl/. Alle URL's in de voetnoten zijn geraadpleegd op 17 april 2023.
[7] Er zijn nog andere mogelijke oorzaken voor discriminatie door AI. Zie voor een overzicht van manieren waarop AI-systemen discriminerend kunnen werken S. Barocas & A.D. Selbst, 'Big Data's disparate impact', *104 California Law Review 671* 2016, https://www.jstor.org/stable/24758720.; F. Zuiderveen Borgesius, *Discrimination, artificial intelligence, and algorithmic decision-making. Report for the European Commission against Racism and Intolerance (ECRI)*, Strasbourg: Council of Europe 2019, www.coe.int, par. III.2. Tilburg Institute for Law, Technology, and Society, *Handreiking voor niet discriminerende algoritmes*, 2021, https://www.tilburguniversity.edu/nl/over/schools/law/departementen/tilt/onderzoek/handreiking.

Of stel dat het HR-personeel van een organisatie die vrouwen heeft gediscrimineerd tijdens sollicitatieprocedures. De organisatie realiseert zich niet dat zijn HR-personeel in het verleden heeft gediscrimineerd. Als de organisatie de HR-beslissingen uit het verleden gebruikt om haar AI-systeem te trainen, zou het AI-systeem die discriminatie kunnen nabootsen. Naar verluidt stuitte een door Amazon ontwikkeld AI-systeem voor de selectie van sollicitanten op een dergelijk probleem. Amazon heeft het systeem overigens niet in de praktijk gebruikt.[9]

AI-systemen kunnen discriminerende beslissingen nemen over sollicitanten en bijvoorbeeld bepaalde etniciteiten benadelen, zelfs als het systeem geen toegang heeft tot gegevens over de etniciteit van mensen. Stel dat een AI-systeem rekening houdt met de postcodes waar sollicitanten wonen. De postcodes zouden kunnen correleren met iemands etniciteit. Het systeem zou dus alle mensen met een bepaalde etniciteit kunnen afwijzen, ook al heeft de organisatie ervoor gezorgd dat het systeem geen rekening houdt met de etniciteit van mensen. AI-systemen kunnen per ongeluk discriminerende effecten hebben: AI-ontwikkelaars en AI-gebruikers realiseren zich mogelijk niet dat het AI-systeem discrimineert.[10]

Stel dat een organisatie wil testen of zijn (nieuwe of bestaande) AI-systeem sollicitanten met een bepaalde etniciteit onterecht discrimineert. Om dit te testen moet de organisatie de etniciteit kennen van zowel de mensen die naar de baan hebben gesolliciteerd, als van de mensen die de organisatie daadwerkelijk heeft aangenomen. Stel dat de helft van de mensen die een sollicitatiebrief hebben gestuurd een immigratie-achtergrond heeft. Het AI-systeem selecteert uit de duizenden brieven de vijftig beste. Een snelle blik suggereert dat het AI-systeem alleen sollicitatiebrieven van witte Nederlanders heeft gekozen. Dergelijke aantallen suggereren dat het AI-systeem moet worden onderzocht op discriminatie. Voor zulk onderzoek zijn gegevens over de etniciteit van de sollicitanten nodig.[11]

---

[8] Zie College voor de Rechten van de Mens, 'Tussenoordeel. De Stichting Vrije Universiteit krijgt de gelegenheid om te bewijzen dat de door haar ingezette antispieksoftware een studente met een donkere huidskleur niet heeft gediscrimineerd', 7 december 2023, https://oordelen.mensenrechten.nl/oordeel/2022-146.
[9] Zie Reuters, 'Amazon scraps secret AI recruiting tool that showed bias against women', https://www.reuters.com/article/us-amazon-com-jobs-automation-insight-idUSKCN1MK08G.
[10] Barocas & Selbst, *California Law Review* 2016/104.

## 3 Non-discriminatiewetgeving

Het recht op non-discriminatie is een mensenrecht.[12] Wij focussen op EU-recht. De EU-wetgeving verbiedt twee vormen van discriminatie: directe en indirecte discriminatie. Volgens de richtlijn inzake rassengelijkheid (over etniciteit) 'wordt onder het beginsel van gelijke behandeling verstaan de afwezigheid van elke vorm van directe of indirecte discriminatie op grond van ras of etnische afstamming.'[13] De Rassenrichtlijn omschrijft *directe* discriminatie als volgt: 'wanneer iemand op grond van ras of etnische afstamming ongunstiger wordt behandeld dan een ander in een vergelijkbare situatie wordt, is of zou worden behandeld.'[14] Een voorbeeld van directe discriminatie is de discriminatie van Zuid-Afrikanen met een donkere huidskleur door het apartheidsregime in Zuid-Afrika in de 20e eeuw.

Directe discriminatie is in de EU-wetgeving verboden. Er zijn enkele specifieke, nauw omschreven uitzonderingen op dit verbod. De Richtlijn inzake rassengelijkheid staat bijvoorbeeld een verschil in behandeling op grond van etniciteit toe als etniciteit 'een wezenlijke en bepalende beroepsvereiste vormt'.[15] Een voorbeeld is de keuze voor een acteur met een donkere huidskleur om de rol van Othello te spelen.[16]

In de non-discriminatiewetgeving worden de gronden zoals etniciteit 'beschermde kenmerken' genoemd. Een AI-systeem dat mensen verschillend behandelt op basis van hun beschermde kenmerken zou rechtstreeks discrimineren. Een hypothetisch voorbeeld van directe discriminatie door een AI-systeem is als de ontwikkelaar het systeem alle vrouwen laat afwijzen.

---

[11] Zie uitgebreider: I. Žliobaitė & B. Custers, 'Using sensitive personal data may be necessary for avoiding discrimination in data-driven decision models', *Artificial Intelligence and Law* 2016/24, afl. 2, p. 183–201, http://link.springer.com/10.1007, doi:10.1007/s10506-016-9182-5.
[12] Artikel 14 EVRM.
[13] Zie artikel 2 Richtlijn 2000/43/EG PB L 180/22.
[14] Artikel 2 lid 2 sub a Richtlijn 2000/43/EG.
[15] Artikel 4 richtlijn 2000/43/EG.
[16] E. Ellis & P. Watson, *EU anti-discrimination law*, Oxford: Oxford University Press 2012.

Ons artikel richt zich op indirecte discriminatie. Van 'indirecte discriminatie' is sprake 'wanneer een ogenschijnlijk neutrale bepaling, maatstaf of handelwijze personen van een bepaald ras of een bepaalde etnische afstamming in vergelijking met andere personen bijzonder benadeelt, tenzij die bepaling, maatstaf of handelwijze objectief wordt gerechtvaardigd door een legitiem doel en de middelen voor het bereiken van dat doel passend en noodzakelijk zijn.'[17]

Er kan sprake zijn van indirecte discriminatie door een AI-systeem als het systeem op het eerste gezicht neutraal is, maar mensen met een beschermd kenmerk blijkt te benadelen. Zelfs als het AI-systeem de beschermde kenmerken negeert, kan het systeem nog steeds discrimineren op basis van neutrale gegevens die blijken te correleren met beschermde kenmerken. De opleiding(en) en universiteit uit het CV van een sollicitant zouden bijvoorbeeld kunnen correleren met etniciteit of een ander beschermd kenmerk.

Het is irrelevant of de organisatie per ongeluk of met opzet discrimineert. Een organisatie is dus altijd verantwoordelijk, zelfs als de organisatie zich niet realiseerde dat zijn AI-systeem indirect discrimineerde. Anders dan voor directe discriminatie, geldt voor indirecte discriminatie een open uitzondering. Indirecte discriminatie is toegestaan als er een objectieve rechtvaardiging bestaat.[18]

### 4    Gegevensbeschermingsrecht

Het recht op privacy en het recht op bescherming van persoonsgegevens zijn beide grondrechten. Privacy wordt bijvoorbeeld beschermd in het Europees Verdrag tot bescherming van de rechten van de mens (1950),[19] en het Handvest van de grondrechten van de Europese Unie (2000).[20] In de EU heeft het recht op bescherming van persoonsgegevens ook de status van een grondrecht.[21]

---

[17] Artikel 2 lid 2 sub b Richtlijn 2000/43/EG.
[18] Artikel 2 lid 2 sub b richtlijn 2000/43/EG.
[19] Artikel 8 van het Europees Verdrag tot bescherming van de rechten van de mens.

Het gegevensbeschermingsrecht verleent rechten aan personen wiens persoonsgegevens worden verwerkt ('betrokkenen'), en legt verplichtingen op aan partijen die persoonsgegevens verwerken. De AVG legt de meeste verantwoordelijkheden bij de 'verwerkingsverantwoordelijke', kort gezegd de partij die het doel van en de middelen voor de verwerking van persoonsgegevens vaststelt.[22] Voor het leesgemak spreken we in dit artikel ook van de 'organisatie'. De AVG heeft niet alleen als doel om privacy te beschermen, maar ook andere grondrechten zoals het recht op non-discriminatie.

## 5 Belemmert de AVG het tegengaan van discriminatie?

### 5.1 Het AVG-verbod op het verwerken van bijzondere persoonsgegevens

De AVG bevat een verbod (met uitzonderingen) op het gebruik van bepaalde soorten extra gevoelige gegevens, de zogenaamde 'bijzondere categorieën van persoonsgegevens'.[23]

De strengere regels voor bijzondere persoonsgegevens beogen onder andere oneerlijke discriminatie te voorkomen. In 1972 zei de Raad van Europa over bijzondere persoonsgegevens: 'In general, information relating to the intimate private life of persons or information which might lead to unfair discrimination should not be recorded or, if recorded, should not be disseminated'.[24] De AVG verwijst in de preambule ook naar het risico van discriminatie, en roept organisaties op te voorkomen dat AI 'discriminerende gevolgen zou hebben'.[25]

---

[20] Artikel 7 van het Handvest van de grondrechten van de Europese Unie.
[21] Artikel 8 van het Handvest van de grondrechten van de Europese Unie.
[22] Artikel 4, lid 7, AVG
[23] Artikel 9 AVG.
[24] Committee of Ministers, Resolution (73)22 on the protection of the privacy of individuals visà-vis electronic data banks in the private sector, 26 September 1973, artikel 1, https://rm.coe.int/1680502830.

Artikel 9, lid 1, AVG verbiedt de verwerking van persoonsgegevens 'waaruit ras of etnische afkomst, politieke opvattingen, religieuze of levensbeschouwelijke overtuigingen, of het lidmaatschap van een vakbond blijken, en verwerking van genetische gegevens, biometrische gegevens met het oog op de unieke identificatie van een persoon, of gegevens over gezondheid, of gegevens met betrekking tot iemands seksueel gedrag of seksuele gerichtheid'.

De meeste beschermde kenmerken in de EU non-discriminatierichtlijnen zijn ook bijzondere persoonsgegevens zoals gedefinieerd in de AVG. Er zijn twee uitzonderingen. Ten eerste zijn 'leeftijd' en 'geslacht' beschermde kenmerken in de non-discriminatiewetgeving, maar geen bijzondere persoonsgegevens in de zin van de AVG.[26] Ten tweede zijn 'politieke opvattingen', 'vakbondslidmaatschap', 'genetische' en 'biometrische' gegevens bijzondere persoonsgegevens in de AVG, maar worden zij niet beschermd door de Europese non-discriminatierichtlijnen. (Overigens verbiedt de Nederlandse Algemene Wet Gelijke Behandeling ook discriminatie op grond van 'politieke gezindheid'.[27])

Figuur 1 toont het onderscheid tussen de 'bijzondere categorieën persoonsgegevens' en de beschermde non-discriminatiegronden.

---

[25] Overweging 71 AVG.
[26] In sommige omstandigheden zouden leeftijd en geslacht bijzondere persoonsgegevens kunnen zijn als ze ruim worden geïnterpreteerd, en worden gezien als 'gezondheidsgegevens', 'biometrische gegevens' of 'genetische gegevens'.
[27] Artikel 1 Awgb.

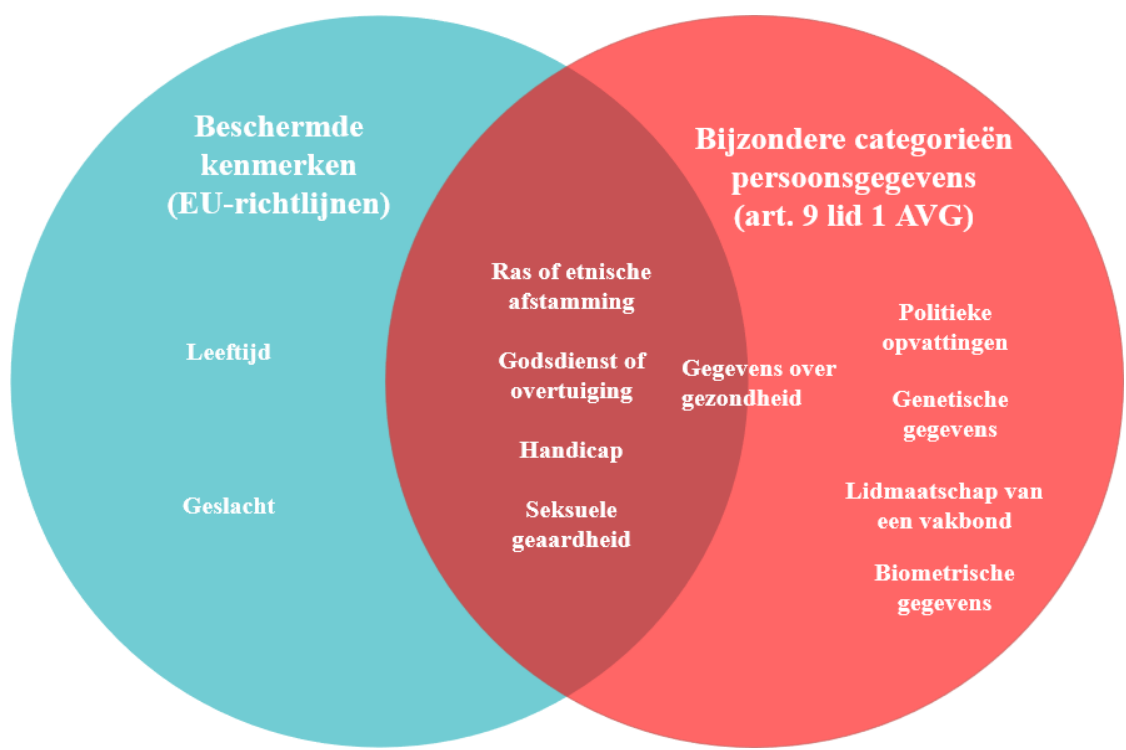

*Figuur 1. De overlap tussen beschermde kenmerken en bijzondere persoonsgegevens.*

Het verbod van de AVG om bijzondere persoonsgegevens te verwerken kan het tegengaan van discriminatie door AI-systemen in de weg staan.[28] Denk aan het volgende scenario. Een organisatie gebruikt een AI-gestuurd wervingssysteem om uit sollicitatiebrieven selecteren. De organisatie wil controleren of zijn AI-systeem per ongeluk tegen bepaalde etnische groepen discrimineert. Daarvoor heeft de organisatie gegevens nodig over de etniciteit van de sollicitanten.

---

[28] K. Alidadi, 'Gauging progress towards equality? Challenges and best practices of equality data collection in the EU', *European Equality Law review* 2017/2, p. 21-22. Y. Al-Zubaidi, 'Some reflections on racial and ethnic statistics for anti-discrimination purposes in Europe', *European Equality Law review*, p. 2020, p. 65.

Daarnaast verbiedt Artikel 9 lid 1 van de AVG het gebruik van gegevens waaruit de etniciteit 'blijkt'. De organisatie mag de etniciteit van zijn sollicitanten dus ook niet afleiden uit andere persoonsgegevens.[29]

### 5.2 De uitzonderingen op het verbod

Artikel 9 lid 2 AVG bevat een lijst met uitzonderingen op het algemene verbod om bijzondere persoonsgegevens te verwerken. Subs a, b, f, g en j bevatten mogelijk relevante uitzonderingen voor het verzamelen van bijzondere categorieën van persoonsgegevens. De uitzonderingen betreffen a) uitdrukkelijke toestemming, en specifieke uitzonderingen voor b) sociale zekerheid, f) rechtsvorderingen, g) redenen van zwaarwegend algemeen belang, en j) onderzoeksdoeleinden. Al deze uitzonderingen moeten restrictief worden uitgelegd.[30]

Sommige non-discriminatierecht-juristen suggereren dat de AVG geen belemmering vormt voor het verzamelen of gebruiken van bijzondere persoonsgegevens om discriminatie te bestrijden.[31] Hierna tonen wij het tegengestelde aan: de AVG verbiedt in de meeste omstandigheden het gebruik van bijzondere persoonsgegevens ter bestrijding van discriminatie.

---

[29] Christopher. Kuner, L.A. Bygrave & Christopher. Docksey, *Commentary on the EU general data protection regulation (GDPR). A commentary. Update of selected articles*, Kettering: Oxford University Press 2021, par. C.1, https://oxford.universitypressscholarship.com/10.1093/oso/9780198826491.001.0001/isbn-9780198826491. Zie ook HvJEU 1 augustus 2022, C-184/20, ECLI:EU:C:2022:601 (OT/Vyriausioji tarnybinės etikos komisija).

[30] Idem C.3.

[31] L. Farkas, *The meaning of racial or ethnic origin in EU law: between stereotypes and identities, European network of legal experts in gender equality and non-discrimination (report for European Commission, Directorate-General for Justice and Consumers)*, Luxembourg: Publications Office of the European Union 2017, https://www.equalitylaw.eu/downloads/4030-the-meaning-of-racial-or-ethinic-origin-in-eu-law-between-stereotypes-and-identities, p. 14. K. Alidadi, 'Gauging progress towards equality? Challenges and best practices of equality data collection in the EU', *European Equality Law review* 2017/2, p. 21-22. Y. Al-Zubaidi, 'Some reflections on racial and ethnic statistics for anti-discrimination purposes in Europe', *European Equality Law review*, p. 2020, p. 22. Zie ook T. Makkonen, *European handbook on equality data: 2016 revision*, LU: Publications Office 2016, https://data.europa.eu/doi/10.2838, p. 27.

### 5.3 Uitdrukkelijke toestemming

Wij bespreken elke mogelijk relevante uitzondering op het verbod om bijzondere gegevens te verwerken achtereenvolgens, te beginnen met toestemming. Het verbod is niet van toepassing als de betrokkene 'uitdrukkelijke toestemming' heeft gegeven voor de verwerking van die persoonsgegevens voor een of meer welbepaalde doeleinden.[32] De vereisten voor geldige toestemming zijn streng.[33] Geldige toestemming vereist dat de toestemming 'specifiek' en 'geïnformeerd' is, en vereist een 'ondubbelzinnige wilsuiting'.[34]

Bovendien schrijft artikel 4 lid 11 AVG voor dat toestemming 'vrij' moet worden gegeven om geldig te zijn. Toestemming van een werknemer (of sollicitant) aan een werkgever is vanwege de onevenredige verhouding tussen beide doorgaans niet vrijwillig in de zin van de AVG.[35]

De vrijwilligheidseis kan het moeilijk maken om AI-gestuurde discriminatie te bestrijden. We keren terug naar ons voorbeeld: een organisatie gebruikt AI om de beste sollicitanten te selecteren en wil zijn AI-systeem controleren op onbedoelde discriminatie. De organisatie zou kunnen overwegen alle sollicitanten om toestemming te vragen om gegevens over hun etniciteit te verzamelen, zodat die informatie kan worden gebruikt om het AI-systeem te controleren. Maar sollicitanten denken misschien dat ze hun kans op een baan verkleinen als ze 'nee' zeggen op een verzoek van de werkgever. Daarom is de toestemming van de sollicitant doorgaans ongeldig.

Misschien kan een systeem worden ontworpen waarbij een sollicitant wel 'vrij' en dus geldig toestemming kan geven. Een organisatie zou bijvoorbeeld alle afgewezen sollicitanten om toestemming kunnen vragen nadat de functie is vervuld. In dat geval zijn sollicitanten misschien niet meer bang dat het weigeren van toestemming hun kansen op de baan verkleint. De organisatie zou het echter ongemakkelijk kunnen vinden om mensen te vragen naar hun etniciteit, religie of seksuele voorkeur. Als te veel mensen weigeren, zal de steekproef bovendien niet representatief zijn.

---

[32] Artikel 9 lid 2 sub a AVG.
[33] Zie artikel 4 lid 11 en artikel 7 AVG.
[34] Artikel 4 lid 11 en artikel 7 AVG.
[35] Zie Overweging 43 van de AVG en European Data Protection Board, *Richtsnoeren 05/2020 inzake toestemming overeenkomstig Verordening 2016/679*, EDPB 2020, https://edpb.europa.eu/sites/default/files/files/file1/edpb_guidelines_202005_consent_nl.pdf, p. 9 en par. 21.

In sommige gevallen, bijvoorbeeld als er geen wanverhouding bestaat tussen de betrokkene en de verantwoordelijke, kunnen organisaties zich wel op toestemming beroepen om het verbod te omzeilen.

### 5.4 Andere uitzonderingen

Hierna bespreken wij kort de andere uitzonderingen op het verbod om bijzondere persoonsgegevens te verwerken, te beginnen met artikel 9 lid 2 sub b AVG:

> (b) de verwerking is noodzakelijk met het oog op de uitvoering van verplichtingen en de uitoefening van specifieke rechten van de verwerkingsverantwoordelijke of de betrokkene op het gebied van het arbeidsrecht en het socialezekerheids- en socialebeschermingsrecht, voor zover zulks is toegestaan bij Unierecht of lidstatelijk recht of bij een collectieve overeenkomst op grond van lidstatelijk recht die passende waarborgen voor de grondrechten en de fundamentele belangen van de betrokkene biedt.[36]

Sub b geldt alleen voor situaties op het gebied van arbeidsrecht en het socialezekerheids- en socialebeschermingsrecht. Een Nederlands voorbeeld is een bepaling in de Uitvoeringswet Algemene verordening gegevensbescherming (UAVG). [37] Grofweg staat de bepaling uit de UAVG toe dat werkgevers gezondheidsgegevens van werknemers verzamelen als dat noodzakelijk is voor bijvoorbeeld hun re-integratie na een ziekte. De Autoriteit Persoonsgegevens staat deze uitzondering alleen toe als dat écht noodzakelijk is.[38]

---

[36] Artikel 9 lid 2 sub b AVG.
[37] Art. 30 UAVG.
[38] Autoriteit Persoonsgegevens, *Besluit tot het opleggen van een bestuurlijke boete,* 2020, https://autoriteitpersoonsgegevens.nl/nl/nieuws/boete-voor-cpa-om-privacyschending-zieke-werknemers.

Sub b zal organisaties die hun AI-systemen willen controleren op discriminatie niet kunnen helpen. Het grootste probleem is dat sub b alleen van toepassing is als de EU-wetgever of de nationale wetgever een specifieke wet heeft aangenomen die het gebruik van bijzondere gegevens mogelijk maakt. Voor zover wij weten heeft geen enkele nationale wetgever in de EU, noch de EU, een wet aangenomen die het gebruik van bijzondere gegevens voor de controle van AI-systemen mogelijk maakt.

De uitzondering in sub f geldt als 'de verwerking noodzakelijk [is] voor de instelling, uitoefening of onderbouwing van een rechtsvordering of wanneer gerechten handelen in het kader van hun rechtsbevoegdheid'. Een organisatie zou kunnen aanvoeren dat hij zijn AI-systemen moet controleren om toekomstige rechtszaken wegens illegale discriminatie te voorkomen. Echter, de uitzondering in sub f alleen geldt voor *concrete* rechtszaken, en is niet gemaakt om eventuele toekomstige rechtszaken te voorkomen.[39] Een organisatie kan deze uitzondering dus meestal niet gebruiken voor het opsporen van discriminatie in haar AI-systemen.

De uitzonderingen g) en j) laten aan de EU en haar lidstaten de ruimte om uitzonderingen in de wet op te nemen voor de verwerking van bijzondere gegevens ter bestrijding van discriminatie.[40] De huidige wetgeving van de EU en Nederland voorziet niet in zo'n uitzondering.[41] Voor de volledigheid vermelden wij dat een organisatie ook moet voldoen aan alle andere vereisten uit de AVG, als hij een uitzondering uit Artikel 9 lid 2 AVG toepast.

Concluderend: de AVG belemmert organisaties die bijzondere persoonsgegevens willen gebruiken om discriminatie door hun AI-systemen tegen te gaan. Soms kan een organisatie het verbod uit de AVG doorbreken door geldige toestemming van de betrokkenen te krijgen. In andere situaties zou een EU- of nationale wet nodig zijn om het gebruik van bijzondere persoonsgegevens voor AI-controle mogelijk te maken. Zou het aannemen van zo'n uitzondering een goed idee zijn?

---

[39] Christopher. Kuner, L.A. Bygrave & Christopher. Docksey, *Commentary on the EU general data protection regulation (GDPR). A commentary. Update of selected articles*, Kettering: Oxford University Press 2021, par. 3, samenvatting, onder 6.
[40] Idem 218.
[41] Het Verenigd Koninkrijk heeft wel een uitzondering. Zie UK Data Protection Act 2018, Schedule 1 Part 2 https://www.legislation.gov.uk/ukpga/2018/12/schedule/1/part/2/crossheading/equality-of-opportunity-or-treatment.

## 6 Een nieuwe uitzondering op het AVG-verbod op het gebruik van bijzondere persoonsgegevens?

### 6.1 Inleiding

De Nederlandse regering heeft overwogen een nationale uitzondering in te voeren op het AVG-verbod om bijzondere persoonsgegeven te gebruiken. In 2020 schreef de Minister van Rechtsbescherming over het verbod in artikel 9 AVG:

> Het kabinet heeft […] aangegeven dat het voornemens is om in afwijking van genoemd verbod, en in bepaalde specifieke gevallen, toe te staan dat bij de ontwikkeling van algoritmische modellen bijzondere persoonsgegevens worden verwerkt, voor zover dat nodig is om [discriminatie] tegen te gaan.[42]

Beleidsmakers buiten Nederland zijn zich bewust van de spanning tussen het beschermen van gegevens en het tegengaan van discriminatie. Het Verenigd Koninkrijk heeft een uitzondering gemaakt op het verbod op het gebruik van bijzondere persoonsgegevens, om discriminatie te bestrijden.[43] Minstens zes landen buiten de EU hebben in hun nationale gegevensbeschermingswet een vergelijkbare uitzondering: Bahrein, Curaçao, Ghana, Jersey, Sint Maarten en Zuid Afrika.[44] De Europese Commissie heeft een voorstel gepubliceerd voor een AI-verordening, met een uitzondering op het verbod uit de AVG, om AI-systemen te controleren op discriminatie (zie paragraaf 7.1).

---

[42] *Kamerstukken II* 2020/21, 26643, nr. 727. Antwoord op vraag 12. Zie ook *Kamerstukken II* 2020/21, 26643, nr. 726, par. 2.2. Zie ook Minister van Justitie en Veiligheid, *Reactie op mededelingen Europese Commissie over de AVG*, 2020, https://www.rijksoverheid.nl/documenten/kamerstukken/2020/12/04/tk-reactie-op-mededelingen-europese-commissie-over-de-avg, p. 3.

[43] UK Data Protection Act 2018, Schedule 1 Part 2 https://www.legislation.gov.uk/ukpga/2018/12/schedule/1/part/2/crossheading/equality-of-opportunity-or-treatment.

[44] Voor een opsomming van deze wetten, zie Van Bekkum & Zuiderveen Borgesius, 'Using sensitive data to prevent discrimination by artificial intelligence: Does the GDPR need a new exception?', Computer Law & Security Review, Volume 48, April 2023, voetnoot 64.

In de volgende paragraaf analyseren wij argumenten voor en tegen het creëren van een uitzondering voor het verzamelen van bijzondere persoonsgegevens ten behoeve van de controle van AI-systemen.

### 6.2 Argumenten voor een uitzondering

Er zijn twee argumenten voor de invoering van een nieuwe uitzondering die het gebruik van bijzondere persoonsgegevens mogelijk maakt om AI-discriminatie tegen te gaan: (i) Organisaties kunnen de gegevens gebruiken om te controleren of een AI-systeem discrimineert, en om discriminatie tegen te gaan. (ii) Het verzamelen van de gegevens heeft een symbolische functie.

(i) Het verzamelen van bijzondere persoonsgegevens is noodzakelijk om discriminatie door AI-systemen tegen te gaan. Organisaties zouden zelf kunnen controleren of hun AI-systeem per ongeluk discrimineert. Soms kunnen organisaties het AI-systeem verbeteren. Zelfs als dat niet mogelijk is, kan een organisatie beslissen het systeem niet meer te gebruiken.

Toezichthouders zouden de AI-systemen van een organisatie gemakkelijker kunnen controleren als die organisaties de etniciteit van al hun werknemers, sollicitanten etc. zouden registreren.[45] Onderzoekers zouden dergelijke gegevens kunnen gebruiken om na te gaan of een AI-systeem discrimineert. Dit argument gaat echter alleen op als een organisatie zijn gegevens met de onderzoeker deelt.

(ii) Een ander soort argument voor het toestaan van zulk gebruik van bijzondere persoonsgegevens is meer symbolisch.[46] Als het bekend is dat een organisatie zijn AI-systemen controleert op discriminatie, dan zouden mensen die organisatie, of AI, meer kunnen vertrouwen.[47] Wij noemen dit symbolische argument voor de volledigheid, maar vinden het geen sterk argument.

---

[45] T. Makkonen, *European handbook on equality data: 2016 revision*, LU: Publications Office 2016, https://data.europa.eu/doi/10.2838, p. 20.
[46] Idem, p. 20.

### 6.3 Argumenten tegen een uitzondering

Er zijn vijf argumenten tegen het aannemen van een uitzondering die het gebruik van bijzondere persoonsgegevens voor het controleren van AI mogelijk maakt.

(i) Mensen kunnen zich ongemakkelijk voelen als gevoelige gegevens over hen worden verzameld of opgeslagen. Mensen kunnen dat gevoel hebben, of die gegevens gebruikt worden of niet.[48] Veel mensen vinden het vervelend als organisaties grote hoeveelheden persoonsgegevens over hen opslaan, zelfs als geen mens ooit naar de gegevens kijkt.

Het Hof van Justitie van de Europese Unie erkent dat alleen al het opslaan van persoonsgegevens een inmenging vormt in het recht op privacy en het recht op bescherming van persoonsgegevens.[49] Ook het Europese Hof voor de Rechten van de Mens erkent dat het opslaan van gevoelige persoonsgegevens het recht op privacy kan schenden, ongeacht de manier waarop die gegevens worden gebruikt.[50]

(ii) Een tweede categorie argumenten betreft de risico's als bijzondere persoonsgegevens worden verzameld en opgeslagen.[51] Zo bestaat het risico op datalekken. Medewerkers van een organisatie of buitenstaanders kunnen onbevoegd toegang krijgen tot de gegevens. Een lek met persoonsgegevens over etniciteit, religie of seksuele voorkeuren kan negatief uitpakken voor de betrokkenen. Vanuit het

---

[47] Zie voor een soortgelijk argument voor het verzamelen van non-discriminatiegegevens in het algemeen K. Alidadi, 'Gauging progress towards equality? Challenges and best practices of equality data collection in the EU', *European Equality Law review* 2017/2, p. 18.
[48] Men kan spreken van subjectieve schade (Calo) of verwachte schade (Gurses). M.R. Calo, 'The Boundaries of Privacy Harm', *86 Indiana Law Journal 31*, p. 1143; F.S. Gurses, *Multilateral Privacy Requirements Analysis in Online Social Network Services*, KU Leuven 2010, p. 312, 87-89.
[49] HvJEU oktober 2013, C 291/12 (*Schwartz/Stadt Bochum*) r.o. 25. Zie ook HvJEU 8 april 2014, C-293/12 en C-594 (*Digital Rights Ireland Ltd*), r.o. 29.
[50] EHRM (Grote Kamer) 25 mei 2021, 58170/13, 62322/14 en 24960/15 (*Big Brother Watch e.a./Verenigd Koninkrijk*), r.o. 392. EHRM 26 maart 1987, 9248/81 (*Leander/Zweden*), r.o. 48.
[51] M.R. Calo, 'The Boundaries of Privacy Harm', *86 Indiana Law Journal 31*, p. 1143; F.S. Gurses, *Multilateral Privacy Requirements Analysis in Online Social Network Services*, KU Leuven 2010, p. 312, 87-89.

oogpunt van gegevensbeveiliging is het beter om zo min mogelijk gevoelige gegevens op te slaan.

Ook bestaat het risico op *function creep*: als zulke gegevens eenmaal zijn opgeslagen, kunnen ze ook voor andere doeleinden gebruikt worden. Als een bedrijf toch al gegevens heeft over de seksuele geaardheid van mensen, kan het die gebruiken voor gepersonaliseerde marketing. En als gegevens eenmaal zijn opgeslagen, kan de politie toegang eisen.

(iii) Organisaties zouden de uitzondering kunnen misbruiken om grote hoeveelheden gevoelige persoonsgegevens te verzamelen, met de bewering dat zij die gegevens nodig hebben om discriminatie te bestrijden. Een te ruime uitzondering zou de deur open kunnen zetten voor grootschalige gegevensverzameling, juist over kwetsbare groepen.[52]

(iv) Het zorgvuldig omgaan met gevoelige gegevens heeft een symbolische functie. Als mensen weten dat een organisatie hun bijzondere persoonsgegevens niet verzamelt, dan zouden zij die organisatie meer kunnen vertrouwen. Wij vinden dit symbolische argument niet sterk.

(v) Het ontwikkelen van niet-discriminerende AI staat in technische zin nog in de kinderschoenen.[53] Zelfs als het gebruik van persoonsgegevens over bijvoorbeeld etniciteit noodzakelijk is om niet-discriminerende AI-systeem te ontwikkelen, betekent dat nog niet dat niet-discriminerende AI ontwikkelen voor een bepaalde taak ook altijd *mogelijk* is. Een organisatie die een AI-systeem ontwikkelt zal zich steeds moeten afvragen of de inzet van het systeem verstandig is.[54] Deze afweging kan veranderen in de toekomst. Hoe beter computerwetenschappers worden in het ontwikkelen van niet-discriminerende AI, des te sterker wordt het argument om het gebruik van bijzondere persoonsgegevens toe te staan.

---

[52] A. Balayn & S. Gürses, *Beyond Debiasing. Regulating AI and its inequalities*, European Digital Rights (EDRi) 2021, https://edri.org/our-work/if-ai-is-the-problem-is-debiasing-the-solution/, p. 94.

[53] M. Andrus e.a., '"What We Can't Measure, We Can't Understand": Challenges to Demographic Data Procurement in the Pursuit of Fairness', *arXiv:2011.02282 [cs]* 2021, p. 257, http://arxiv.org/abs/2011.02282. K. Holstein e.a., 'Improving fairness in machine learning systems: What do industry practitioners need?', *Proceedings of the 2019 CHI Conference on Human Factors in Computing Systems* 2019, http://arxiv.org/abs/1812.05239.

[54] Zie Tilburg Institute for Law, Technology, and Society & Ministerie van Binnenlandse Zaken en Koninkrijksrelaties, Handreiking non-discriminatie Artificial Intelligence (AI), 2022, https://www.rijksoverheid.nl/documenten/rapporten/2022/12/05/handreiking-non-discriminatie-artificial-intelligence-ai, p. 21.

Al met al zijn er verschillende argumenten voor en tegen het aannemen van een uitzondering die het gebruik van bijzondere persoonsgegevens mogelijk maakt om AI-gedreven discriminatie tegen te gaan. Als een uitzondering zou worden aangenomen, zou die ook waarborgen moeten bevatten om de risico's te beperken.

## 7   Mogelijke waarborgen als een uitzondering wordt aangenomen

### 7.1  Waarborgen in de voorgestelde AI-verordening

Een voorstel van de EU illustreert enkele mogelijkheden voor waarborgen. Begin 2021 presenteerde de Europese Commissie een voorstel voor een verordening omtrent AI-systemen, met daarin een uitzondering op het verbod op het gebruik van bijzondere persoonsgegevens. De voorgestelde uitzondering is als volgt geformuleerd:

> Voor zover dit strikt noodzakelijk is om de monitoring, opsporing en correctie van vertekeningen te waarborgen in verband met de AI-systemen met een hoog risico, mogen de aanbieders van dergelijke systemen bijzondere categorieën persoonsgegevens, zoals bedoeld in artikel 9, lid 1, van [de AVG] verwerken, mits passende waarborgen worden geboden voor de grondrechten en fundamentele vrijheden van natuurlijke personen, met inbegrip van technische beperkingen voor het hergebruik en het gebruik van ultramoderne beveiligings- en privacybeschermende maatregelen, zoals pseudonimisering of versleuteling wanneer anonimisering aanzienlijke gevolgen kan hebben voor het nagestreefde doel.[55]

---

[55] Artikel 10 lid 5 AI-Verordening https://eur-lex.europa.eu/legal-content/EN/TXT/?uri=CELEX%3A52021PC0206.

De voorgestelde bepaling bevat verschillende waarborgen om misbruik van de gegevens van bijzondere aard te voorkomen.

### 7.1.1 Strikt noodzakelijk om discriminatie tegen te gaan

De uitzondering uit de AI-verordening geldt alleen 'voor zover dit strikt noodzakelijk is'. De uitdrukking 'strikt noodzakelijk' stelt een zwaardere eis dan alleen 'noodzakelijk'. Uit de rechtspraak van het HvJEU blijkt dat het woord 'noodzakelijk' al streng moet worden uitgelegd, in het voordeel van de betrokkene.[56] Organisaties mogen dus alleen een beroep doen op de uitzondering in de AI-verordening als het gebruik van bijzondere persoonsgegevens echt noodzakelijk is.

### 7.1.2 De uitzondering van de AI-verordening geldt alleen voor aanbieders van AI-systemen met een hoog risico

De uitzondering van de AI-verordening geldt alleen voor AI-systemen met een hoog risico. AI-systemen met een hoog risico kunnen in twee soorten worden onderverdeeld: Ten eerste, producten die al onder bepaalde EU-wetgeving over gezondheid en veiligheid vallen, zoals medische hulpmiddelen. Ten tweede, acht soorten AI-systemen die in een bijlage bij de AI-verordening worden genoemd.[57] Als de wetgever een nieuwe uitzondering wil maken op het gebruik van bijzondere persoonsgegevens, dan zou hij de uitzondering kunnen beperken tot, bijvoorbeeld, AI-systemen die ernstige discriminatierisico's met zich brengen.

### 7.1.3 Passende waarborgen

De uitzondering in de voorgestelde AI-verordening zegt dat bijzondere persoonsgegevens kunnen worden gebruikt om discriminatie op grond van AI tegen te gaan, 'mits passende waarborgen worden geboden voor de grondrechten en fundamentele vrijheden van natuurlijke personen'.[58] De bepaling geeft voorbeelden van dergelijke waarborgen: 'technische beperkingen voor het hergebruik en het gebruik van ultramoderne beveiligings- en privacybeschermende maatregelen […].' [59] Als een uitzondering zou worden aangenomen om het gebruik van bijzondere persoonsgegevens voor de bestrijding van AI-discriminatie mogelijk te maken, zou een vergelijkbare eis moeten worden opgenomen.

---

[56] HvJEU mei 2017, C-13/16 (*Rigas*), r.o. 30. HvJEU 8 april 2014, C-293/12 en C-594 (*Digital Rights Ireland Ltd*), r.o. 52. EHRM 25 maart 1983, 5947/72; 6205/73; 7052/75; 7061/75; 7107/75; 7113/75; 7136/75 (*Silver e.a./Verenigd Koninkrijk*), r.o. 97.
[57] AI-verordening, Annex III.

Sommige elementen in de voorgestelde uitzondering van de AI-verordening zijn omstreden. Zo laat de voorgestelde uitzondering in het midden wie bepaalt wat de passende waarborgen zijn. Met de huidige tekst lijkt de AI-ontwikkelaar zelf te mogen beslissen. Het Europees Parlement overweegt strengere en expliciet genoemde waarborgen toe te voegen, zoals: het gebruik van synthetische of anonieme datasets, pseudonimisering, waarborgen voor beveiliging van de gegevens, waarborgen voor toegangscontrole tot de data, een verbod op het delen van de data met derde partijen, en de data alleen bewaren voor zolang dat strikt noodzakelijk is.[60]

### 7.2 Andere mogelijke waarborgen

Zijn andere waarborgen mogelijk? Misschien biedt een synthetische, anonieme dataset ('*synthetic data*') een oplossing, op basis van de originele persoonsgegevens. Het idee is dat de *synthetic data* dezelfde verdeling van mensen vertegenwoordigen, maar dat de data niet meer aan individuen kan worden gekoppeld. De belofte is dat dergelijke gegevens veilig kunnen worden gebruikt, en de AVG niet meer van toepassing is. Er moeten nog steeds bijzondere persoonsgegevens worden verzameld om de synthetische dataset te maken, maar de bijzondere gegevens hoeven minder lang bewaard te worden.

---

[58] Artikel 10 lid 3 AI-verordening.
[59] Artikel 10, lid 3, AI-verordening.
[60] Europees Parlement, *DRAFT Compromise Amendments on the Draft Report Proposal for a regulation of the European Parliament and of the Council on harmonised rules on Artificial Intelligence (Artificial Intelligence Act) and amending certain Union Legislative Acts*, COM(2021)0206 – C9 0146/2021 – 2021/0106(COD), p. 58 https://www.europarl.europa.eu/meetdocs/2014_2019/plmrep/COMMITTEES/CJ40/DV/2023/05-11/ConsolidatedCA_IMCOLIBE_AI_ACT_EN.pdf .

Het is omstreden hoe effectief het gebruik van *synthetic data* is voor het testen van AI-systemen. Bovendien bieden synthetic data volgens verschillende auteurs amper betere privacybescherming dan bestaande oudere anonymiseringstechnieken. [61]

Andere mogelijke waarborgen zijn meer organisatorisch. Een derde partij zou bijvoorbeeld de bijzondere persoonsgegevens kunnen verzamelen, opslaan en gebruiken voor de controle van het AI-systeem. De organisatie die het AI-systeem gebruikt of ontwikkelt, hoeft de gegevens dan niet meer zelf op te slaan. Zo'n aanpak roept veel vragen op. Welke partij zou die gegevens moeten opslaan? Een mogelijkheid is misschien het Centraal Bureau voor de Statistiek. Of misschien kan een toezichthouder betrouwbare onderzoekers aanwijzen en hen toegang geven tot de bijzondere persoonsgegevens om discriminatie door AI-systemen tegen te gaan. De financiële en praktische haalbaarheid van zulke oplossingen vergt meer onderzoek.[62]

## 8    Conclusie

Dit artikel liet zien dat de AVG een verbod bevat op het gebruik van bijzondere persoonsgegevens om AI-systemen op discriminatie te kunnen controleren. In sommige gevallen kunnen mensen het verbod opzijzetten door toestemming te geven voor het gebruik van hun bijzondere persoonsgegevens. Aan de andere kant is de toestemming in veel gevallen niet geldig. Andere AVG-uitzonderingen op het verbod vereisen een specifieke nationale of EU-rechtelijke bepaling die voldoet aan de juiste waarborgen. Er zijn in Nederland en de EU geen wettelijke uitzonderingen die het gebruik van bijzondere persoonsgegevens mogelijk maken voor het controleren van AI-systemen. Kortom, de AVG-regels over bijzondere persoonsgegevens staan het tegengaan van discriminatie door AI in de weg.

---

[61] T. Stadler, B. Oprisanu en C. Troncoso, 'Synthetic Data - Anonymisation Groundhog Day', *31st USENIX security symposium* 2022, https://www.usenix.org/conference/usenixsecurity22/presentation/stadler. Zie ook S.M. Bellovin, P.K. Dutta & N. Reitinger, 'Privacy and Synthetic Datasets', *22 Stan. Tech. L. Rev. 1* 2019, p. 14 en verder, https://law.stanford.edu/wp-content/uploads/2019/01/Bellovin_20190129.pdf.

[62] Zie ook N. Kilbertus e.a., 'Blind Justice: Fairness with Encrypted Sensitive Attributes', *arXiv:1806.03281 [cs, stat]* 2018, p. 1-2, http://arxiv.org/abs/1806.03281.

Er is een spanning tussen het gegevensbeschermingsrecht en het voorkomen van discriminatie. Deze moet niet gezien worden als een spanning tussen privacybelangen en non-discriminatiebelangen. Een van de doelen van de strenge AVG-regels voor bijzondere persoonsgegevens is het tegengaan van discriminatie. Databanken met gevoelige gegevens kunnen misbruikt worden om te discrimineren. Er is dus reden om voorzichtig te zijn met het introduceren van een uitzondering op het verbod op het verzamelen van bijzondere categorieën persoonsgegevens - zelfs als die uitzondering tot doel heeft om discriminatie te voorkomen.

Heeft de AVG een nieuwe uitzondering nodig op het bijzondere-persoonsgegevens-verbod, om het mogelijk te maken AI te controleren op discriminatie? We brachten de argumenten voor en tegen zo'n nieuwe uitzondering in kaart, maar zullen die in deze conclusie niet herhalen. Voor beide standpunten valt veel te zeggen. Als al een nieuwe uitzondering aangenomen wordt, dan zou de wet strenge waarborgen moeten bevatten om de daarmee samenhangende risico's tegen te gaan.

Academici zoals wij zouden niet moeten beslissen hoe het evenwicht tussen de verschillende belangen moet worden gevonden. Zo'n beslissing moet democratisch gelegitimeerd zijn. Maar academici kunnen wel proberen het debat te informeren. We hopen dat dit artikel daarbij helpt.

* * *